\documentclass[prl,onecolumn,showpacs,superscriptaddress]{revtex4}

\usepackage{amsmath}
\usepackage{amsfonts}
\usepackage{amssymb}
\usepackage{graphicx}
\usepackage{bm}

\begin{document}

\title{Quantum complementarity and logical indeterminacy}

\author{{\v C}aslav Brukner}
\affiliation{Institute for Quantum Optics and Quantum Information,
Austrian Academy of Sciences, Boltzmanngasse 3, A-1090 Vienna,
Austria} \affiliation{Faculty of Physics, University of Vienna,
Boltzmanngasse 5, A-1090 Vienna, Austria}

\date{\today}

\begin{abstract}

Whenever a mathematical proposition to be proved requires more
information than it is contained in an axiomatic system, it can
neither be proved nor disproved, i.e. \. it is undecidable, or
logically undetermined, within this axiomatic system. I will show
that certain mathematical propositions on a $d$-valent function of a
binary argument can be encoded in $d$-dimensional quantum states of
mutually unbiased basis (MUB) sets, and truth values of the
propositions can be tested in MUB measurements. I will then show
that a proposition is undecidable within the system of axioms
encoded in the state, if and only if the measurement associated with
the proposition gives completely random outcomes.

\end{abstract}


\maketitle

The theorems of Bell~\cite{BELL}, Kochen and Specker~\cite{KS} as
well as of Greenberger, Horne and Zeilinger~\cite{GHZ} showed that
the mere concept of co-existence of local elements of physical
reality is in a contradiction with quantum mechanical predictions.
Apart from the known loopholes, which are considered by the majority
of physicists to be of technical nature, all experiments confirmed
the quantum predictions. This implies that either the assumption of
``elements of reality'', or ``locality'' or both must fail. Most
working scientists seem to hold fast to the concept of ``elements of
reality''. One of the reasons for this tendency might be that it is
not clear how to base a theory without this concept. While one
should leave all options open, it should be noted that maintaining
the assumption of realism and denying locality faces certain
conceptual problems~\cite{peres}. But, perhaps more importantly in
my view is that so far this approach could not encourage any new
phenomenology that might result in the hope for a progressive
research program.

An alternative to this is to arrive at a new understanding of
probabilities which is not based on our ignorance about some
prederminated properties. What is then the origin of probabilities?
What makes the probabilities different at all? Here I will show that
certain mathematical propositions on a $d$-valent function of a
binary argument can be encoded in $d$-dimensional quantum states
(qudits), and truth values of the propositions can be tested in
corresponding quantum measurements. I will then show that logically
independent propositions correspond to measurements in mutually
unbiased basis (MUB) sets. In quantum theory, a pair of orthonormal
bases $\{|k\rangle \},0\le k \le d-1$ and $\{|l\rangle \}, 0\le l
\le d-1$ in a Hilbert space $\mathbb{C}^d$ are said to be unbiased
if the modulus square of the inner product between any basis vector
from $\{|k\rangle \}$ with any other basis vector from $\{|l \rangle
\}$ satisfies $ |\langle k | l \rangle|^2 = 1/d$. A set of bases for
which each pair of bases are unbiased is said to be mutually
unbiased~\cite{MUB}. If one assumes that there is a fundamental
limit on how much information a quantum system can carry (``a single
qudit carries one dit of information''), and that this information
is exhausted in defining one of the propositions (taken as an
axiom), then the measurements that correspond to logically
independent propositions must give irreducibly random outcomes. This
allows to derive the probabilities ($=1/d$) for outcomes of the MUB
measurements without directly invoking quantum theory, but by
looking if the proposition is definite or ``undecidable'' within the
axiomatic set.

In 1982, Chaitin gave an information theoretical formulation of
mathematical undecidability suggesting that it arises whenever a
proposition to be proven and the axioms contain together more
information than the set of axioms
alone~\cite{chaitin,caludejuergensen}. In this work, when relating
mathematical undecidability to quantum randomness, I will
exclusively refer to the incompleteness in Chaitin's sense and not
to the original work of G\"{o}del. Furthermore, I will consider
mathematical undecidability in those axiomatic systems which can be
completed and which therefore are {\it not} subject to G\"{o}del's
incompleteness theorem~\cite{fusnote1}.

Consider a $d$-valent function $f(x)\in {0,...,d-1}$ of a single
binary argument $x\in\{0,1\}$, with $d$ a prime
number~\cite{footnote}. There are $d^2$ such functions. We will
partition the functions into $d+1$ different ways following the
procedure of Ref.~\cite{paterekdakicbrukner}. In a given partition,
the $d^2$ functions will be divided into $d$ different groups each
containing $d$ functions. Enumerating the first $d$ partitions by
the integer $a = 0,...,d-1$ and the groups by $b = 0,...,d-1$, the
groups of functions are generated from the formula:
\begin{equation}
f(1) = a f(0) \oplus b, \label{PRIME_QUESTION}
\end{equation}
where the sum is modulo $d$. In the last partition, enumerated by
$a=d$, the functions are divided into groups $b = 0,...,d-1$
according to the functional value $f(0)=b$. The functions can be
represented in a table in which $a$ enumerates the rows of the
table, while $b$ enumerates different columns. For all but the last
row the table is built in the following way : (i) choose the row,
$a$, and the column, $b$; (ii) vary $f(0) = 0,...,d-1$ and compute
$f(1)$ according to Eq.~(\ref{PRIME_QUESTION}); (iii) write pairs
$f(0)\,f(1)$ in the cell. The last row ($a=d$) is built as follows:
(i) choose the column $b$; (ii) vary $f(1) =0,...,d-1$ and put
$f(0)=b$; (iii) write pairs $f(0)\,f(1)$ in the cell.  For example,
for $d=3$, one has
\begin{equation}
\begin{tabular}[b]{ccc|ccc|ccc}
\multicolumn{3}{c|}{$b=0$} & \multicolumn{3}{c|}{$b=1$} & \multicolumn{3}{c}{$b=2$} \\
\hline  \hline $00$ & $10$ & $20$ & $01$& $11$& $21$ & $02$& $12$&
$22$
\\ \hline
$00$ & $11$ & $22$ & $01$& $12$& $20$ & $02$& $10$& $21$
\\ \hline
$00$ & $12$ & $21$ & $01$& $10$& $22$ & $02$& $11$& $20$
\\ \hline
 $00$ & $01$ & $02$ & $10$& $11$& $12$ & $20$& $21$&
$22$
\\ \hline
\hline
\end{tabular}
 \quad
\begin{tabular}[b]{c}
\hline  \hline ``$f(1) = b$''
\\ \hline
``$f(1) = f(0) \oplus  b$''
\\ \hline
``$f(1) = 2 f(0) \oplus b$''
\\ \hline
``$f(0) = b$''
\\ \hline
 \hline
\end{tabular}
\label{3DESIGN}
\end{equation}

All groups (cells in the table) of functions that do not belong to
the last row are specified by the proposition:
\begin{eqnarray}
\{a,b\}: \mbox{``The function values $f(0)$ and $f(1)$ satisfy $f(1)
= a f(0) \oplus b$''}, \label{proposition}
\end{eqnarray}
while those from the last row $(a=d)$ by
\begin{eqnarray}
\{d,b\}: \mbox{``The function value $f(0)=b$''}.
\label{proposition1}
\end{eqnarray}
The propositions corresponding to different partitions $a$ are {\it
independent} from each other.  For example, if one postulates the
proposition (A) ``$f(1) = a f(0) \oplus b$'' to be true, i.e. if we
choose it as an ``axiom'', then it is possible to prove that
``theorem'' (T1) ``$f(1) = a f(0) \oplus b'$'' is false for all
$b'\neq b$. Proposition (T1) is decidable within the axiom (A).
Within the same axiom (A) it is, however, impossible to prove or
disprove ``theorem'' (T2) ``$f(1) = m f(0) \oplus n$'' with $ m\neq
a$. Having only axiom (A), i.e.\ only {\it one} dit of information,
there is not enough information to know also the truth value of
(T2). Ascribing truth values to two propositions belonging to two
different partitions, e.g. to both (A) and (T2), would require {\it
two} dits of information. Hence, in Chaitin's sense, proposition
(T2) is mathematically {\it undecidable} within the system
containing the single axiom (A).

So far, we have made only logical statements. To make a bridge to
physics consider a hypothetical device -- "preparation device" --
that can encode a mathematical axiom $\{a,b\}$  of the
type~(\ref{proposition}) or (\ref{proposition1}) into a property of
a physical system by setting a "control switch" of the apparatus in
a certain position $\{a,b\}$. In an operational sense the choice of
the mathematical axiom is entirely defined by the switch position as
illustrated in Figure 1 (top). We make no particular assumptions on
the physical theory (e.g., classical or quantum) that underlies the
behavior of the system, besides that it {\it fundamentally limits
the information content of the system to one dit of information}.
Furthermore, we assume that there is a second device -- a
"measurement apparatus" -- that can test the truth value of a chosen
mathematical proposition again by setting a control switch of the
apparatus to a certain position associated to the proposition. The
choice of the switch position $\{m\}$, $m \in\{0,...,d\}$,
corresponds to a performance of one of the $d+1$ possible
measurements on the system and the occurrence of a $d$-valued
outcome $n$ in the measurement is identified with finding
proposition $\{m,n\}$ (of the type~(\ref{proposition}) or
(\ref{proposition1})) being true. Consider now a situation where the
preparation device is set on $\{a,b\}$, while the measurement
apparatus on $\{m\}$. If $m=a$, the outcome confirms the axiom,
i.e.\ one has $n=b$. This is why we say that measurement $\{m\}$
tests mathematical proposition $\{a,b\}$. What will be the outcome
in a single run of the experiment if $m\neq a$?

I will show that devices from the previous paragraph are not
hypothetical at all. In fact, they can be realized in quantum
mechanics. The argument is entirely based on
Ref.~\cite{paterekdakicbrukner}. In the basis of generalized Pauli
operator $\hat Z$, denoted as $| \kappa \rangle$, $k \in
\{0,...,d-1\}$, we define two elementary operators
\begin{equation}
\hat Z | \kappa \rangle = \eta_d^{\kappa} | \kappa \rangle, \qquad
\hat X | \kappa \rangle = | \kappa + 1 \rangle,
\end{equation}
where $\eta_d = \exp{(i 2 \pi/d)}$ is a complex $d$th root of unity.
The eigenstates of  the $\hat X \hat Z^a$ operator, $a \in
\{0,...,d-1\}$, expressed in the $\hat Z$ basis, are given by $| j
\rangle_a = (1/\sqrt{d}) \sum_{\kappa=0}^{d-1} \eta_d^{-j \kappa - a
s_{\kappa}} | \kappa \rangle$, where $s_{\kappa} =
\kappa+...+(d-1)$~\cite{UNITARY_MUB}, and the $\hat Z$ operator
shifts them: $\hat Z | j \rangle_a = | j - 1 \rangle_a$. To encode
the axiom $\{a,b\}$ into a quantum state the preparation device is
set to prepare state $|0\rangle_a$ and then to apply the unitary
$\hat U = \hat X^{f(0)} \hat Z^{f(1)}$ on it (Figure 1, down). The
action of the device is, for $a=0,...,d-1$ and up to a global phase,
$\hat U \propto (\hat X \hat Z^a)^{f(0)} \hat Z^b$, which follows
from Eq.~(\ref{PRIME_QUESTION}) and the commutation relation for the
elementary operators, $\hat Z \hat X = \eta_d \hat X \hat Z$. The
state leaving the preparation device is shifted exactly $b$ times
resulting in $|-b\rangle_a$. For the case $a=d$ the state is
prepared in the eigenstate $|0\rangle_d\equiv|0\rangle$ of the
operator $\hat Z$ and the unitary transforms it into, up to the
phase factor, $|+b\rangle_d$. When the switch of the measurement
apparatus is set to $\{m\}$ it measures the incoming state in the
basis $\{|0\rangle_m,...,|d-1\rangle_m\}$. For $m=a$ the measurement
will confirm the axiom $\{a,b\}$ giving outcome $b$. In all other
cases, the result will be completely {\it random}. This follows from
the fact that the eigenbases of $\hat X \hat Z^a$ for $a=0,...,d-1$
$(\hat{Z}^0\equiv \mathbf{1})$ and eigenbasis of $\hat Z$ are known
to form a complete set of $d+1$ mutually unbiased basis
sets~\cite{UNITARY_MUB}. They have the property that a system
prepared in a state from one of the bases will give completely
random results if measured in any other basis, i.e. $|_a\langle
b|n\rangle_m|^2=1/d$ for all  $b$, $n$ and $a \neq m$. The previous
discussion suggests, however, that probabilities for MUB
measurements can be justified without directly invoking quantum
theory, by looking if the proposition is definite, or undecidable,
within the axiomatic system. For the analysis of logical
propositions and MUB measurements on composite separable and
entangled quantum systems see Ref.~\cite{patereketal}.

\begin{figure}
\begin{center}
\includegraphics[width=12cm]{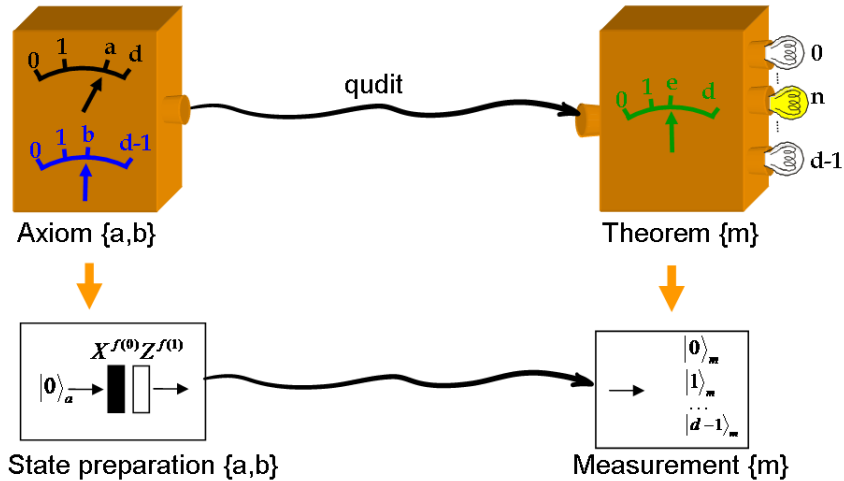}
\end{center}
\caption{Quantum experiment testing (un)decidability of mathematical
propositions (\ref{proposition}) and (\ref{proposition1}). A qudit
($d$-dimensional quantum state) is initialized in a definite quantum
state $|0\rangle_a$ of one of $d+1$ mutually unbiased bases sets
$a\in \{0,...,d\}$. Subsequently, the unitary transformation $
\hat{U}=\hat{X}^{f(0)}\hat{Z}^{f(1)}$ which encodes the $d$-valued
function with functional values $f(0)$ and $f(1)$ is applied to the
qudit. The final state encodes the proposition: ``$f(1) = a f(0)
\oplus b$'' for $a=0,...,d-1$ or the proposition:``$f(0)=b$'' for
$a=d$. The measurement apparatus is set to measure in the $m$-th
basis $\{|0\rangle_m,..., |d-1\rangle_m\}$, which belongs to one of
$d+1$ mutually unbiased basis sets $m\in\{0,...,d\}$. It tests the
propositions: ``$f(1) = m f(0) \oplus n$'' for $m=0,...,d-1$ or
``$f(0)=n$'' for $m=d$.} \label{figure}
\end{figure}

Most working scientists hold fast to the viewpoint according to
which randomness can only arise due to the observer's ignorance
about predetermined well-defined properties of physical systems. But
the theorems of Kochen and Specker~\cite{KS} and Bell~\cite{BELL}
have seriously put  such a belief in question. I argue that an
alternative viewpoint according to which quantum randomness is
irreducible is vindicable. As proposed by Zeilinger~\cite{zeilinger}
an individual quantum system contains only a limited information
content (``a single qudit carries one dit of information''). I have
shown here that one can encode a finite set of axioms in a quantum
state and test the truth values of mathematical propositions in
quantum measurements. If the proposition is decidable within the
axiomatic system, the outcome of the measurement will be definite.
However, if it is undecidable, the response of the system must not
contain any information whatsoever about the truth value of the
undecidable proposition, and it cannot ``refuse'' to give an
answer~\cite{footnote1}. Unexplained and perhaps unexplainable, it
inevitably gives an outcome -- a "click" in a detector or a flash of
a lamp -- whenever measured. I suggest that the individual outcome
must then be irreducible random, reconciling mathematical
undecidability with the fact that a system always gives an
``answer'' when ``asked'' in an experiment.

I am grateful to T. Paterek, R. Prevedel, J. Kofler, P. Klimek, M.
Aspelmeyer and A. Zeilinger for numerous discussions on the topic.
This work is based on Ref.~\cite{paterekdakicbrukner} and
\cite{patereketal}. I acknowledge financial support from the
Austrian Science Fund (FWF), the Doctoral Program CoQuS and the
European Commission under the Integrated Project Qubit Applications
(QAP).



\begin{thebibliography}{99}

\bibitem{BELL} J. Bell, Physics {\bf 1}, 195 (1964).
\bibitem{KS} S. Kochen and E. Specker, J. Math. and
Mech. \textbf{17}, 59 (1967).
\bibitem{GHZ} D. Greenberger, M. A. Horne, and A. Zeilinger,
in:~\textit{Bell's Theorem, Quantum Theory, and Conceptions of the
Universe}, ed.\ M. Kafatos (Kluwer Academic Publishers, 1989);
electronic version: arXiv:0712.0921v1 [quant-ph].
\bibitem{peres} A. Peres, {\it Quantum Theory: Concepts and Methods}
(Kluwer Academic, Dordrecht, 1995).
\bibitem{MUB} I. D. Ivanovic, J. Phys. A \textbf{14}, 3241 (1981). W. K. Wooters and B. D. Fields, Ann. Phys. (N.Y.) \textbf{191}, 363 (1989).
\bibitem{chaitin} G. J. Chaitin, Int. J. Theor. Phys. {\bf 21}, 941-954 (1982).
\bibitem{caludejuergensen} C. S. Calude and H. J\"{u}rgensen, Appl. Math. {\bf 35}, 1-15 (2005).
\bibitem{fusnote1} For a possible relation between G\"{o}del's theorem and physics see J. D. Barrow, G\"{o}del and Physics, arXiv:physics/0612253
(2006); paper delivered at ``Horizons of Truth'', Kurt G\"{o}edel
Centenary Meeting, Vienna, 27-29th April 2006. For a relation
between algorithmic randomness and quantum indeterminacy see M. A.
Stay and C. S. Calude, Int. J. Theor. Phys. {\bf 44}, 1053-1065
(2005); K. Svozil, Phys. Lett. A {\bf 143}, 433-437 (1990); and C.
S. Calude, and K. Svozil, arXiv:quant-ph/0611029.
\bibitem{footnote} The considerations here can be generalized to all dimensions that are
powers of primes. This is related to the fact that in quantum theory
in these cases a complete set of mutually unbiased bases is known to
exit. In all other cases this is an open question and goes beyond
the scope of this paper (see, for example,
Ref.~\cite{paterekdakicbrukner}).
\bibitem{paterekdakicbrukner} T. Paterek, B. Daki\'{c} and {\v C}. Brukner, Phys.
Rev. A \textbf{79}, 012109 (2009).
\bibitem{UNITARY_MUB} S. Bandyopadhyay
\emph{et al}., Algorithmica {\bf 34}, 512 (2002).
\bibitem{zeilinger} A. Zeilinger, Found. Phys. {\bf 29},
631-643 (1999).
\bibitem{footnote1} To put it in a grotesque way the system is not allowed to response ``I am undecidable, I cannot give an answer.''
\bibitem{patereketal} T. Paterek, R. Prevedel, J. Kofler, P. Klimek, M. Aspelmeyer, A.
Zeilinger, and {\v C}. Brukner, electronic version: arXiv:0811.4542.

\end{thebibliography}
\end{document}